\documentclass[a4paper,final,12pt,onecolumn]{article}%
\usepackage{amsfonts}
\usepackage{amsmath}
\usepackage{amssymb}
\usepackage{graphicx}
\usepackage{version}
\usepackage{accents}%
\providecommand{\U}[1]{\protect\rule{.1in}{.1in}}

\begin{document}

\title{On the direct quantization of Maxwell field}
\author{{\small W. Benarab}$^{1}${\small , Z. Belhadi}$^{2\ast}$\\$^{1,2}${\small Laboratoire de physique th\'{e}orique, Facult\'{e} des
sciences exactes,}\\{\small Universit\'{e} de Bejaia. 06000 Bejaia, Alg\'{e}rie.}\\zahir.belhadi@univ-bejaia.dz\\warda.benarab@univ-bejaia.dz }
\date{}
\maketitle

\begin{abstract}
In this paper, we apply the generalized integration constants method \cite{z3}
in field theory to quantize Maxwell and the Klein-Gordon free fields. The
study is performed in both position and momentum spaces, to obtain equal-time
Dirac brackets among the fields and their conjugate momenta. The idea is to
obtain the brackets near the initial instant using the Taylor polynomial
expansion, and then deduce directly their expressions at any later time. In
the case of the Maxwell field, the interdependence of the field components
(constraints) requires the use the Helmholtz theorem to separate the
transversal and longitudinal parts.


\end{abstract}

\section{\textbf{ }INTRODUCTION}

\qquad Canonical quantization is based on the canonical formalism of classical
mechanics where Poisson brackets play a fundamental role to determine the
commutators of different quantum operators through the help of the
correspondence principle. However, this approach works only in regular cases,
hence the need to introduce the Dirac brackets in singular cases,
characterized by the presence of constraints. Two standard approaches provide
the same results: Dirac formalism \cite{1} and Faddeev-Jackiw method \cite{2}.

The method of integration constants "CI" is another way to obtain the Dirac
brackets in the case of exactly solvable singular systems \cite{z1,z2}. This
completely different approach, is straightforward, highly accessible and does
not require advanced mathematical tools . Its inconvenient\ is the necessity
to have access to the equations of motion general solution with all the
independent integration constants, in order to use them in computing
fundamental variables brackets.

Recently, the CI approach was improved in order to include non integrable
singular systems \cite{z3}. The idea is to use the first-order Taylor
expansion near the initial conditions as a replacement for the general
solution and calculate different brackets at the initial time. To deduce these
brackets at any later instant, one can use their time covariance. Unlike the
other approaches, this generalized CI method does not distinguish between the
constraints, because they are all satisfied by the general solution or by the
Taylor expansion. Several systems with finite degree of freedom and in the
field theory are successfully treated using this method \cite{z1,z2,z3}.

In the following, we will use the generalized "CI" method in the field theory
framework to quantize the Klein-Gordon scalar field and the Maxwell vectorial
field. Our goal is to show clearly how this novel approach operates through
these two examples in both ordinary and Fourier spaces.

\subsection{The generalized "CI" method}

\qquad Consider a classical system described by an autonomous singular
Lagrangian $L(q,\dot{q})$ where the $q=(q_{1},...,q_{N})$ and the $\dot
{q}=(\dot{q}_{1},...,\dot{q}_{N})$ are respectively the generalized
coordinates and the generalized velocities. Suppose that the equations of
Euler-Lagrange can be solved and the general solution can be expressed as
$q(t)=Q(t,C)$ and $p(t)=\frac{\partial L}{\partial\dot{q}}(t,C)=P(t,C)$ and
contains all independent integration constants $C=(C_{1},C_{2},...,C_{M}).$
Here, $M\leqslant2N$ because the presence of constraints reduces the number of
these constants. The Hamiltonian\ is a conserved quantity and has an
expression of the form $H=H(Q(t,C),P(t,C))=H(C).$

The CI method \cite{z1,z2} uses the Hamilton equations $\dot{q}_{i}%
=\{q_{i},H\}$ and $\dot{p}_{i}=\{p_{i},H\}$ expressed in term of the
integration constants to calculate the brackets $\{C_{I},C_{J}\}\ $using the
equations $\frac{\partial Q_{i}(t,C)}{\partial t}=\{Q_{i}(t,C),H(C)\}$ and
$\frac{\partial P_{i}(t,C)}{\partial t}=\{P_{i}(t,C),H(C)\}.$ After expanding,
we arrive at the fundamental relations of the CI method \
\begin{align}
\frac{\partial Q_{i}(t,C)}{\partial t}  &  =\{C_{I},C_{J}\}\frac{\partial
Q_{i}}{\partial C_{I}}\frac{\partial H}{\partial C_{J}}\\
\frac{\partial P_{i}(t,C)}{\partial t}  &  =\{C_{I},C_{J}\}\frac{\partial
P_{i}}{\partial C_{I}}\frac{\partial H}{\partial C_{J}}.
\end{align}
At this point, the brackets $\{C_{I},C_{J}\}$ are obtained directly by
identification. Then they will be used to determine the brackets of the
fundamental variables $q=Q(t,C)$ and $p=P(t,C)$. \bigskip

One of the obstacles encountered with the initial version of the "CI" method
is the difficulty of finding the general solution of non-integrable systems.
Fortunately, a generalization of this approach can be done using a Taylor
expansion of the solution near of the initial conditions \cite{z3}. Indeed, if
$\xi(t)=(q(t),p(t)),$ the Taylor formula to the first order will be%
\begin{equation}
\xi_{I}(t)=\tilde{\xi}_{I}+\left.  \dot{\xi}_{I}\right\vert _{t=0}\text{
}t+O(t^{2}) \label{1539}%
\end{equation}
where the $\tilde{\xi}_{I}$ are the initial conditions $\xi_{I}(0)=\tilde{\xi
}_{I}$.$\ $The derivatives at the initial instant, $\left.  \dot{\xi}%
_{I}\right\vert _{t=0}$ are obtained from the equations of motion
(Euler-Lagrange equations) as a function of $\tilde{\xi}_{I}$. The Hamiltonian
of the system being conserved, it will also be expressed as a function of the
initial conditions $H=H(\tilde{\xi})=\tilde{H}.$ To access to the
brackets$\{\tilde{\xi}_{I},\tilde{\xi}_{J}\}$, we have to use the result
(\ref{1539}) and impose the equations of Hamilton in the initial instant
\begin{equation}
\left.  \frac{d\xi_{I}}{dt}\right\vert _{t=0}=\left.  \{\xi_{I},H(\xi
)\}\right\vert _{t=0}\Rightarrow\left.  \dot{\xi}_{I}\right\vert
_{t=0}=\{\tilde{\xi}_{I},\tilde{\xi}_{J}\}\frac{\partial\tilde{H}}%
{\partial\tilde{\xi}_{J}}.
\end{equation}
Immediately, the brackets $\{\tilde{\xi}_{I},\tilde{\xi}_{J}\}$ are obtained
by a direct identification of the members of previous equalities. To get the
brackets $\{\xi_{I}(t),\xi_{J}(t)\}$ at anylater time, we take advantage of
their temporal covariance \cite{z3}, as follows :
\begin{equation}
\{\tilde{\xi}_{I},\tilde{\xi}_{J}\}=\Theta_{IJ}(\tilde{\xi})\text{
}\Rightarrow\text{ }\{\xi_{I}(t),\xi_{J}(t)\}=\Theta_{IJ}(\xi(t)).
\end{equation}
In other words, the bracket $\{\xi_{I},\xi_{J}\}$ keeps its form invariant
during time evolution. This means that by substituting the argument
$\tilde{\xi}$ with the argument $\xi(t)$ in the function $\Theta_{IJ}$, we can
derive the corresponding bracket.

To summarize, by using the Hamilton equations in the neighborhood of the
initial instant, the temporal evolution of dynamical variables enables us to
compute the brackets of a system at this instant and then extend them to the future.

\section{Klein-Gordon real field}

\subsection{Klein-Gordon field in the position space}

\qquad The Lagrangian density of the real Klein--Gordon field $\phi(\vec
{x},t)$ is of the form $L=\int dx^{3}(\ \frac{1}{2}\partial_{\mu}\phi
\partial^{\mu}\phi-\frac{m^{2}}{2}\phi^{2})$ where the mass $m>0.$ Using the
definition of the canonical momentum $\pi=\frac{\partial%
\mathcal{L}%
}{\partial(\partial_{t}\phi)},$ and the Euler-Lagrange equations
$\frac{\partial%
\mathcal{L}%
}{\partial\phi}-\partial_{\beta}\frac{\partial%
\mathcal{L}%
}{\partial(\partial_{\beta}\phi)}=0$, one find the Klein-Gordon field
equation
\begin{equation}
\partial_{\mu}\partial^{\mu}\phi=m^{2}\phi\Leftrightarrow\left\{
\begin{array}
[c]{l}%
\dot{\phi}=\pi\\
\dot{\pi}=\Delta\phi-m^{2}\phi
\end{array}
\right.  \label{1}%
\end{equation}
and the canonical Hamiltonian
\begin{equation}
H=\int\frac{dx^{3}}{2}\left(  \pi^{2}+\partial_{i}\phi\text{ }\partial_{i}%
\phi+m^{2}\phi^{2}\right)  .
\end{equation}
\bigskip

Now, let's consider a first-order Taylor expansion of the equation (\ref{1})
solution, starting from the initial conditions $\left.  \phi(\vec
{x},t)\right\vert _{t=0}\ =\Phi(\vec{x})$ and $\left.  \pi(\vec{x}%
,t)\right\vert _{t=0}=\Pi(\vec{x}).$ We obtain easily the relations%
\begin{align}
\phi(\vec{x},t)  &  =\Phi(\vec{x})+\Pi(\vec{x})\text{ }t+O(t^{2})\label{4}\\
\pi(\vec{x},t)  &  =\Pi(\vec{x})+\left(  \Delta\Phi(\vec{x})-m^{2}\Phi(\vec
{x})\right)  \text{ }t+O(t^{2}) \label{3}%
\end{align}
The Hamiltonian is conserved in time ($H=\left.  H\right\vert _{t=0})$,
therefore%
\begin{equation}
H=\int\frac{dx^{3}}{2}\left(  \Pi(\vec{x})^{2}+\partial_{i}\Phi(\vec{x})\text{
}\partial_{i}\Phi(\vec{x})+m^{2}\Phi(\vec{x})^{2}\right)  \label{2}%
\end{equation}
We are now going to use the Hamilton equations at initial instant$\left.
\dot{\phi}(\vec{x},t)\right\vert _{t=0}=\left\{  \left.  \phi(\vec
{x},t)\right\vert _{t=0},H\right\}  $ and $\left.  \dot{\pi}(\vec
{x},t)\right\vert _{t=0}=\left\{  \left.  \pi(\vec{x},t)\right\vert
_{t=0},H\right\}  $, to calculate the brackets of $\Phi(\vec{x})$ and
$\Pi(\vec{x}).$ Directly from the equations (\ref{4}) et (\ref{3}), we deduce
the following relations%
\begin{align}
\Pi(\vec{x})  &  =\left\{  \Phi(\vec{x}),H\right\} \label{qis}\\
\Delta\Phi(\vec{x})-m^{2}\Phi(\vec{x})  &  =\left\{  \Pi(\vec{x}),H\right\}
\label{qisqis}%
\end{align}
Now, using the expression (\ref{2}) of the Hamiltonian, one can write
\begin{equation}
\Pi(\vec{x})=\int dy^{3}\left(  \left\{  \Phi(\vec{x}),\Pi(\vec{y})\right\}
\Pi(\vec{y})-\left\{  \Phi(\vec{x}),\Phi(\vec{y})\right\}  \Delta^{\prime}%
\Phi(\vec{y})+m^{2}\left\{  \Phi(\vec{x}),\Phi(\vec{y})\right\}  \Phi(\vec
{y})\right)  \label{45}%
\end{equation}
where $\partial_{i}^{\prime}=\frac{\partial}{\partial y^{i}}.$ In this
expression, we replaced the term $\left\{  \Phi(\vec{x}),\partial_{i}^{\prime
}\Phi(\vec{y})\right\}  \partial_{i}^{\prime}\Phi(\vec{y})$ by

\noindent$\partial_{i}^{\prime}\left[  \left\{  \Phi(\vec{x}),\Phi(\vec
{y})\right\}  \partial_{i}^{\prime}\Phi(\vec{y})\right]  -\left\{  \Phi
(\vec{x}),\Phi(\vec{y})\right\}  \partial_{i}^{\prime}\partial_{i}^{\prime
}\Phi(\vec{y}),$ then we omitted the divergence term $\partial_{i}^{\prime
}\left[  \left\{  \Phi(\vec{x}),\Phi(\vec{y})\right\}  \partial_{i}^{\prime
}\Phi(\vec{y})\right]  .$ The initial conditions are totally independent,
therefore after the identification between the left and right sides of
(\ref{45}) we get
\begin{equation}
\left\{  \Phi(\vec{x}),\Pi(\vec{y})\right\}  =\delta(\vec{x}-\vec{y})\text{
\ \ \ et \ \ \ }\left\{  \Phi(\vec{x}),\Phi(\vec{y})\right\}  =0.
\end{equation}
In the same way, the equation (\ref{3}) gives the relation%
\[
\Delta\Phi(\vec{x})-m^{2}\Phi(\vec{x})^{2}=\int dy^{3}\left[  \left\{
\Pi(\vec{x}),\Pi(\vec{y})\right\}  \Pi(\vec{y})+\left\{  \Pi(\vec{x}%
),\Phi(\vec{y})\right\}  \mathbf{(-}\Delta\Phi(\vec{x})+m^{2}\Phi(\vec{x}%
)^{2})\right]
\]
which allow to determine the brackets%
\begin{equation}
\left\{  \Pi(\vec{x}),\Phi(\vec{y})\right\}  =-\delta(\vec{x}-\vec{y})\text{
\ \ \ et \ \ \ }\left\{  \Pi(\vec{x}),\Pi(\vec{y})\right\}  =0.
\end{equation}
As the brackets remain covariant over time, we finally deduce that
\begin{equation}
\left\{  \phi(t,\vec{x}),\pi(t,\vec{y})\right\}  =\delta(\vec{x}-\vec
{y})\text{ \ \ \ et \ \ \ }\left\{  \phi(t,\vec{x}),\phi(t,\vec{y})\right\}
=\left\{  \pi(t,\vec{x}),\pi(t,\vec{y})\right\}  =0 \label{opop}%
\end{equation}
These are the well-known brackets of the real scalar field of Klein-Gordon
theory \cite{5}.

\subsection{Klein-Gordon field in the momentum space}

\qquad In this section, we use the relations (\ref{qis}) and (\ref{qisqis})
and the Fourier transforms of the real Klein--Gordon field and its conjugate
momentum at the initial instant to find the well-known commutation relations.

Any complex number $z$ can be put in the form $z=\frac{\alpha+\beta}{2}%
+i\frac{\alpha-\beta}{2}$ where $\left(  \alpha,\beta\right)  \in%
\mathbb{R}
^{2}.$ Now, if $z=z(\vec{k})$, where $\vec{k}\in%
\mathbb{R}
^{3}$ and $z(\vec{k})=z(-\vec{k})^{\ast}$ then
\begin{equation}
z(\vec{k})=\frac{1+i}{2}\text{ }\alpha(\vec{k})+\frac{1-i}{2}\text{ }%
\alpha(-\vec{k}).
\end{equation}
Using the Fourier transform, the initial condition $\Phi(\vec{x})$ can be
written as $\Phi(\vec{x})=\int f(\vec{k})$ $e^{i\vec{x}\cdot\vec{k}}dk^{3}$,
where $f(\vec{k})=f(-\vec{k})^{\ast}$ to ensure its reality$.$ So there is a
real function $\alpha(\vec{k})$ such as $f(\vec{k})=\frac{1+i}{2}\alpha
(\vec{k})+\frac{1-i}{2}\alpha(-\vec{k})$ and the previous Fourier transform
becomes%
\begin{equation}
\Phi(\vec{x})=\int dk^{3}\alpha(\vec{k})\left(  \frac{1+i}{2}e^{i\vec{x}%
\cdot\vec{k}}+\frac{1-i}{2}e^{-i\vec{x}\cdot\vec{k}}\right)  . \label{753}%
\end{equation}
The same situation occurred with the initial condition $\Pi(\vec{x})$ which
takes the form
\begin{equation}
\Pi(\vec{x})=\int dk^{3}\beta(\vec{k})\left(  \frac{1+i}{2}e^{i\vec{x}%
\cdot\vec{k}}+\frac{1-i}{2}e^{-i\vec{x}\cdot\vec{k}}\right)  . \label{456321}%
\end{equation}
One can express the Hamiltonian at the initial instant $H$ in term of the
function $\alpha$ and $\beta$ as%
\begin{equation}
H=\int\frac{(2\pi)^{3}}{2}dq^{3}\left(  \beta(\vec{q})^{2}+(m^{2}+\vec{q}%
^{2})\alpha(\vec{q})^{2}\right)  . \label{951}%
\end{equation}

Now, let's use the expressions (\ref{753}) et (\ref{951}) to get
\begin{align}
\left\{  \Phi(\vec{x}),H\right\}   &  =(2\pi)^{3}\int dk^{3}dq^{3}%
{\Huge [}\left\{  \alpha(\vec{k}),\beta(\vec{q})\right\}  \beta(\vec
{q})\left(  \frac{1+i}{2}e^{i\vec{x}\cdot\vec{k}}+\frac{1-i}{2}e^{-i\vec
{x}\cdot\vec{k}}\right) \nonumber\\
&  \text{ \ \ \ \ }+\left\{  \alpha(\vec{k}),\alpha(\vec{q})\right\}
(m^{2}+\vec{q}^{2})\text{ }\alpha(\vec{q})\left(  \frac{1+i}{2}e^{i\vec
{x}\cdot\vec{k}}+\frac{1-i}{2}e^{-i\vec{x}\cdot\vec{k}}\right)  {\Huge ]}
\label{dfr}%
\end{align}
Taking account the relations (\ref{qis}) and (\ref{456321}), we arrive at the
commutation relations
\begin{equation}
\left\{  \alpha(\vec{k}),\beta(\vec{q})\right\}  =\frac{1}{(2\pi)^{3}}%
\delta(\vec{k}-\vec{q})\text{ \ \ \ et \ \ \ \ }\left\{  \alpha(\vec
{k}),\alpha(\vec{q})\right\}  =0 \label{res}%
\end{equation}
We also have%
\begin{align}
\left\{  \Pi(\vec{x}),H\right\}   &  =(2\pi)^{3}\int dk^{3}dq^{3}%
{\Huge [}\left\{  \beta(\vec{k}),\beta(\vec{q})\right\}  \beta(\vec{q})\left(
\frac{1+i}{2}e^{i\vec{x}\cdot\vec{k}}+\frac{1-i}{2}e^{-i\vec{x}\cdot\vec{k}%
}\right) \nonumber\\
&  \text{ \ \ \ \ }+\left\{  \beta(\vec{k}),\alpha(\vec{q})\right\}
(m^{2}+\vec{q}^{2})\text{ }\alpha(\vec{q})\left(  \frac{1+i}{2}e^{i\vec
{x}\cdot\vec{k}}+\frac{1-i}{2}e^{-i\vec{x}\cdot\vec{k}}\right)  {\Huge ]}%
\end{align}
and%
\begin{equation}
\Delta\Phi(\vec{x})-m^{2}\Phi(\vec{x})=-\int dk^{3}(m^{2}+\vec{k}^{2})\text{
}\alpha(\vec{k})\left(  \frac{1+i}{2}e^{i\vec{x}\cdot\vec{k}}+\frac{1-i}%
{2}e^{-i\vec{x}\cdot\vec{k}}\right)  .
\end{equation}
As $\Delta\Phi(\vec{x})-m^{2}\Phi(\vec{x})=\left\{  \Pi(\vec{x}),H\right\}  $
(see (\ref{qisqis})), we directly deduce the relations%
\begin{equation}
\left\{  \beta(\vec{k}),\alpha(\vec{q})\right\}  =-\frac{1}{(2\pi)^{3}}%
\delta^{3}(\vec{k}-\vec{q})\text{ \ \ \ et \ \ \ \ }\left\{  \beta(\vec
{k}),\beta(\vec{q})\right\}  =0 \label{ase}%
\end{equation}
Finally, equations (\ref{res}) and (\ref{ase}) and expressions (\ref{753}) and
(\ref{456321}) enable us to derive the brackets
\begin{equation}
\left\{  \Phi(\vec{x}),\Pi(\vec{y})\right\}  =\delta(\vec{x}-\vec{y})\text{
\ \ \ \ \ \ \ }\left\{  \Phi(\vec{x}),\Phi(\vec{y})\right\}  =\text{\ }%
\left\{  \Pi(\vec{x}),\Pi(\vec{y})\right\}  =0
\end{equation}
\ \qquad\qquad\qquad\qquad\qquad

Thus, we have obtained the same result already found in the previous section
(\ref{opop}). For the second time, we were able to reproduce the correct
commutation relations of the Klein-Gordon field with the help of the
generalized integration constants method.

\section{Maxwell field}

\section{Maxwell field in position space}

The electromagnetic field is not trivial as it has constraints that are
directly related to the gauge symmetry. This field is described by the free
Lagrangian
\begin{equation}
L=-\frac{1}{4}\int dx^{3}F_{\mu\nu}F^{\mu\nu}%
\end{equation}
where $F_{\mu\nu}$ is the electromagnetic tensor defined in term of the
four-vector $A_{\mu}$ as $F_{\mu\nu}=\partial_{\mu}A_{\nu}-\partial_{\nu
}A_{\mu}$ where $\mu,\nu\in\{1,2,3,4\}$. In the Coulomb gauge $\partial
_{i}A^{i}=0$ and $A_{0}=0,$ the equations of motion $\partial_{\beta}%
F^{\beta\lambda}=0$ and the canonical momenta $\pi^{\mu}=-F^{0\lambda}$ take
the reduced form%
\begin{align}
A_{0}  &  =0\text{ \ \ \ \ and \ \ \ \ }\pi_{0}=0\label{a}\\
\partial_{i}A^{i}  &  =0\text{ \ \ \ \ and\ \ \ \ }\dot{A}^{i}=-\pi
^{i}\label{b}\\
\partial_{i}\pi^{i}  &  =0\text{ \ \ \ \ and \ \ \ \ }\dot{\pi}^{i}=-\Delta
A^{i} \label{c}%
\end{align}
The Hamiltonian of this system is then given by%
\begin{equation}
H=\frac{1}{2}\int dx^{3}\left(  \left(  \pi^{i}\right)  ^{2}+\left(
\partial_{j}A^{i}\right)  ^{2}-\partial_{i}A^{j}\partial_{j}A^{i}\right)
\label{d}%
\end{equation}

Supposing that, the initial conditions are $A^{i}(\vec{x},0)=$ $\Lambda
^{i}(\vec{x})$ and $\pi^{i}(\vec{x},0)$ $=\Pi^{i}(\vec{x}),$ the equations
(\ref{b}) and (\ref{c}) taken at the initial instant ($t=0$), allow us to get
the relations $\partial_{i}\Lambda^{i}=0$, $\partial_{i}\Pi^{i}=0,$ $\left.
\dot{A}^{i}\right\vert _{t=0}=-\Pi^{i}$ and $\left.  \dot{\pi}^{i}\right\vert
_{t=0}=-\Delta\Lambda^{i}.$ Now, one can deduce the Taylor expansion to the
first-order of the electromagnetic field
\begin{align}
A^{i}(\vec{x},t)  &  =\Lambda^{i}(\vec{x})-\Pi^{i}(\vec{x})\text{ }%
t+O(t^{2})\label{11}\\
\pi^{i}(\vec{x},t)  &  =\Pi^{i}(\vec{x})-\Delta\Lambda^{i}(\vec{x})\text{
}t+O(t^{2})\label{22}\\
\partial_{i}\Lambda^{i}(\vec{x})  &  =0\text{ \ \ \ \ \ \ and \ \ \ \ \ \ }%
\partial_{i}\Pi^{i}(\vec{x})=0 \label{33}%
\end{align}
and the Hamiltonian becomes%
\begin{equation}
H=\int\left(  \frac{1}{2}\Pi^{i}(\vec{x})\Pi^{i}(\vec{x})+\frac{1}{2}%
\partial_{j}\Lambda^{i}(\vec{x})\partial_{j}\Lambda^{i}(\vec{x})-\frac{1}%
{2}\partial_{i}\Lambda^{j}(\vec{x})\partial_{j}\Lambda^{i}(\vec{x})\right)
dx^{3} \label{486}%
\end{equation}
At this stage, we impose the equations of Hamilton at the initial instant
$\left.  \dot{A}^{k}(\vec{x},t)\right\vert _{t=0}=\left\{  \left.  A^{k}%
(\vec{x},t)\right\vert _{t=0},H\right\}  $ and $\left.  \dot{\pi}^{k}(\vec
{x},t)\right\vert _{t=0}=\left\{  \left.  \pi^{k}(\vec{x},t)\right\vert
_{t=0},H\right\}  $ to obtain the different brackets$.$ From (\ref{11}),
(\ref{22}) and (\ref{33}), it results
\begin{align}
-\Pi^{i}(\vec{x})  &  =\left\{  \Lambda^{i}(\vec{x}),H\right\}  \text{
\ \ with\ \ \ }\partial_{i}\Pi^{i}(\vec{x})=0\label{jij}\\
-\Delta\Lambda^{i}(\vec{x})  &  =\left\{  \Pi^{i}(\vec{x}),H\right\}  \text{
\ \ with\ \ \ }\partial_{i}\Lambda^{i}(\vec{x})=0 \label{ijii}%
\end{align}
Inserting (\ref{486}) in (\ref{jij}) to arrive to the expanded form
\begin{align}
-\mathbf{\Pi}^{i}\mathbf{(\vec{x})}  &  =\int dy^{3}{\huge (}\left\{
\Lambda^{i}(\vec{x}),\Pi^{j}(\vec{y})\right\}  \mathbf{\Pi}^{j}\mathbf{(\vec
{y})}+\left\{  \Lambda^{i}(\vec{x}),\partial_{j}^{\prime}\Lambda^{k}(\vec
{y})\right\}  \partial_{j}^{\prime}\Lambda^{k}(\vec{y})\nonumber\\
&  \text{ \ \ \ \ \ \ \ \ \ \ }-\frac{1}{2}\left\{  \Lambda^{i}(\vec
{x}),\partial_{k}^{\prime}\Lambda^{j}(\vec{y})\right\}  \partial_{j}^{\prime
}\Lambda^{k}(\vec{y})-\frac{1}{2}\left\{  \Lambda^{i}(\vec{x}),\partial
_{j}^{\prime}\Lambda^{k}(\vec{y})\right\}  \partial_{k}^{\prime}\Lambda
^{j}(\vec{y}){\huge )}%
\end{align}
As the $\Lambda^{i}$ and $\Pi^{i}$ are completely independent, the
contribution of right terms that do not contain $\Pi^{i}$ must be canceled.
This is possible if the barckets $\left\{  \Lambda^{k}(\vec{x}),\Lambda
^{i}(\vec{y})\right\}  $ are zero
\begin{equation}
\left\{  \Lambda^{k}(\vec{x}),\Lambda^{i}(\vec{y})\right\}  =0.
\end{equation}
The equation becomes
\begin{equation}
\Pi_{i}(\vec{x})=-\int dy^{3}\Pi_{j}(\vec{y})\text{ }\left\{  \Lambda_{i}%
(\vec{x}),\Pi_{j}(\vec{y})\right\}  \text{ \ \ \ \ \ with\ \ \ \ \ }%
\partial_{i}\Pi^{i}(\vec{x})=0 \label{458}%
\end{equation}

It is evident that identification can not proceed in this manner because the
components $\Pi^{i}$ are related by the relation $\partial_{i}\Pi^{i}=0,$ this
is why we appeal to the Helmholtz theorem \cite{6,7,8,9}, particularly
applicable in this situation. Indeed, any vectorial field $V_{i}(\vec{x})$
physically zero at infinity, decomposes in a unique way as $V_{i}(\vec
{x})=V_{i}^{\bot}(\vec{x})+V_{i}^{//}(\vec{x})$ where $V_{i}^{\bot}(\vec{x})$
and $V_{i}^{//}(\vec{x})$ designate respectively the transversal and the
longitudinal components of the field $V_{i}(\vec{x})\ $satisfying the
conditions $\operatorname{div}\vec{V}^{\bot}=0$ and $\operatorname{curl}%
\vec{V}^{//}=\vec{0}.$ These components are given by $V_{i}^{\bot}(\vec
{x})=\int dy^{3}\delta_{ij}^{\bot}(\vec{x}-\vec{y})$ $V_{j}(\vec{y})$ and
$V_{i}^{//}(\vec{x})=\int dy^{3}\delta_{ij}^{//}(\vec{x}-\vec{y})$ $V_{j}%
(\vec{y})$ where we recognize the transverse delta $\delta_{ij}^{\bot}(\vec
{x}-\vec{y})$ and the longitudinal delta $\delta_{ij}^{//}(\vec{x}-\vec{y})$
functions$.$ Our field $\Pi^{i}$ verifies the gauge condition
$\operatorname{div}\vec{\Pi}=0,$ so we deduce that $\Pi_{i}(\vec{x})=\Pi
_{i}^{\bot}(\vec{x})$ and $\Pi_{i}^{//}=0.$ Explicitly,
\begin{equation}
\Pi_{i}(\vec{x})=\int dy^{3}\delta_{ij}^{\bot}(\vec{x}-\vec{y})\Pi_{j}(\vec
{y})\text{ \ \ \ \ \ \ with \ \ \ \ \ }\partial_{i}\Pi_{i}(\vec{x})=0
\label{452}%
\end{equation}
where
\begin{equation}
\delta_{ij}^{\bot}(\vec{x}-\vec{y})=\delta_{ij}\text{ }\delta(\vec{x}-\vec
{y})+\partial_{i}\partial_{j}\left(  \frac{1}{4\pi|\vec{x}-\vec{y}|}\right)  .
\end{equation}
Now, by comparing between (\ref{458}) and (\ref{452}), we directly obtain the
desired bracket%
\begin{equation}
\{\Lambda_{i}(\vec{x}),\Pi_{^{j}}(\vec{y})\}=-\delta_{ij}^{\bot}(\vec{x}%
-\vec{y})
\end{equation}
\qquad\qquad

Let's try to repeat the same steps with the equation (\ref{ijii}). We have
\begin{align*}
-\Delta\Lambda^{i}(\vec{x})  &  =\int dy^{3}(\left\{  \Pi^{i}(\vec{x}),\Pi
^{j}(\vec{y})\right\}  \Pi^{j}(\vec{y})+\left\{  \Pi^{i}(\vec{x}),\partial
_{j}^{\prime}\Lambda^{k}(\vec{y})\right\}  \partial_{j}^{\prime}\Lambda
^{k}(\vec{y})\\
&  -\frac{1}{2}\left\{  \Pi^{i}(\vec{x}),\partial_{k}^{\prime}\Lambda^{j}%
(\vec{y})\right\}  \partial_{j}^{\prime}\Lambda^{k}(\vec{y})-\frac{1}%
{2}\left\{  \Pi^{i}(\vec{x}),\partial_{j}^{\prime}\Lambda^{k}(\vec
{y})\right\}  \partial_{k}^{\prime}\Lambda^{j}(\vec{y}))
\end{align*}
Using the properties of partial derivatives and the relation $\partial
_{j}^{\prime}\Lambda^{j}(\vec{y})=0,$ the previous relation reduces to the
form
\begin{equation}
\mathbf{\Delta\Lambda}^{i}\mathbf{(\vec{x})}=\int dy^{3}\left(  -\left\{
\Pi^{i}(\vec{x}),\Pi^{j}(\vec{y})\right\}  \Pi^{j}(\vec{y})+\left\{  \Pi
^{i}(\vec{x}),\Lambda^{k}(\vec{y})\right\}  \mathbf{\Delta}^{\prime
}\mathbf{\Lambda}^{k}\mathbf{(\vec{y})}\right)
\end{equation}
The fields $\Lambda^{i}$ and $\Pi^{j}$ are completely independent, so by
identification%
\begin{equation}
\left\{  \Pi^{i}(\vec{x}),\Pi^{j}(\vec{y})\right\}  =0
\end{equation}
and the resulting equation is%
\begin{equation}
\Delta\Lambda_{i}(\vec{x})=\int dy^{3}\left\{  \Pi_{i}(\vec{x}),\Lambda
_{j}(\vec{y})\right\}  \text{ }\Delta^{\prime}\Lambda_{j}(\vec{y})\text{
\ \ \ \ with \ \ \ \ \ }\partial_{i}\Lambda_{i}=0.
\end{equation}
On the other hand, the vector field $\Delta\vec{\Lambda}$ is physically zero
at infinity and it is transversal because $\operatorname{div}\Delta
\vec{\Lambda}=\Delta\operatorname{div}\vec{\Lambda}=0,$ then we deduce that
\begin{equation}
\Delta\Lambda_{i}(\vec{x})=\int dy^{3}\delta_{ij}^{\bot}(\vec{x}-\vec
{y})\text{ }\Delta^{\prime}\Lambda_{j}(\vec{y})\text{ \ \ \ \ with
\ \ \ \ }\partial_{i}\Lambda^{i}=0.
\end{equation}
A direct identification between the two previous equations will allow us to
have the bracket%
\begin{equation}
\left\{  \Pi_{i}(\vec{x}),\Lambda_{j}(\vec{y})\right\}  =\delta_{ij}^{\bot
}(\vec{x}-\vec{y})
\end{equation}
These brackets conserve their form in time
\begin{equation}
\{A_{i}(\vec{x},t),\pi_{^{j}}(\vec{y},t)\}=-\delta_{ij}^{\bot}(\vec{x}-\vec
{y})
\end{equation}

We have, therefore, obtained the well-known brackets of the canonical
quantization of the Maxwell field \cite{5,6}.

\section{Maxwell field in the momentum space}

Using Fourier transform, the initial conditions $\vec{A}(\vec{x}%
,0)=\vec{\Lambda}(\vec{x})$ and $\vec{\pi}(\vec{x},0)=\vec{\Pi}(\vec{x})$ of
the Maxwell field and its conjugate monenta can be written as
\begin{align}
\vec{\Lambda}(\vec{x})  &  =\int dk^{3}\vec{\alpha}(\vec{k})\left(  \frac
{1+i}{2}\text{ }e^{i\vec{x}\cdot\vec{k}}+\frac{1-i}{2}\text{ }e^{-i\vec
{x}\cdot\vec{k}}\right) \label{12369874}\\
\vec{\Pi}(\vec{x})  &  =\int dk^{3}\vec{\beta}(\vec{k})\left(  \frac{1+i}%
{2}\text{ }e^{i\vec{x}\cdot\vec{k}}+\frac{1-i}{2}\text{ }e^{-i\vec{x}\cdot
\vec{k}}\right)  \label{pln}%
\end{align}
where $\vec{\alpha}(\vec{k})$ and $\vec{\beta}(\vec{k})$ are real vectorial
functions. The Coulombian gauge $\partial_{i}\Lambda_{i}=0$ implies the
relation $\vec{\alpha}(\vec{k})\cdot\vec{k}=0,\forall$ $\vec{k}\in R^{3}$ (
this means that the vector $\vec{k}$ is perpendicularly to the plan containing
the vector $\vec{\alpha}(\vec{k})$)$.$ One can always find a real basis
$\left\{  \vec{\varepsilon}_{1}(\vec{k}),\vec{\varepsilon}_{2}(\vec
{k})\right\}  $ for this plan verifying the equations
\begin{align}
\vec{\varepsilon}_{\lambda}(\vec{k})\cdot\vec{\varepsilon}_{\lambda^{\prime}%
}(\vec{k})  &  =\delta_{\lambda\lambda^{\prime}}\text{ \ \ \ },\text{
\ \ \ }\vec{\varepsilon}_{1}(\vec{k})\wedge\vec{\varepsilon}_{2}(\vec
{k})=\frac{\vec{k}}{\left\vert \vec{k}\right\vert },\\
\vec{\varepsilon}_{1}(-\vec{k})  &  =\vec{\varepsilon}_{1}(\vec{k})\text{
\ \ \ },\text{ \ \ \ }\vec{\varepsilon}_{2}(-\vec{k})=-\vec{\varepsilon}%
_{2}(\vec{k})
\end{align}
The three vectors $\left\{  \varepsilon_{1}(\vec{k}),\varepsilon_{2}(\vec
{k}),\frac{\vec{k}}{\left\vert \vec{k}\right\vert }\right\}  $ form a basis of
all space, and they respect the closure relation
\begin{equation}
\sum_{\lambda=1}^{2}\left(  \vec{\varepsilon}_{\lambda}(\vec{k})\right)
_{i}\left(  \vec{\varepsilon}_{\lambda}(\vec{k})\right)  _{j}=\delta
_{ij}-\frac{k_{i}k_{j}}{\left\vert k^{2}\right\vert } \label{sss}%
\end{equation}
Now, the vector $\vec{\alpha}(\vec{k})$ can be expressed as $%
{\displaystyle\sum\limits_{\lambda=1}^{2}}
\alpha_{\lambda}(\vec{k})\ \vec{\varepsilon}_{\lambda}(\vec{k})$ where
$\alpha_{1}(\vec{k})$ and $\alpha_{2}(\vec{k})$ are real components and
(\ref{12369874}) became
\begin{equation}
\vec{\Lambda}(\vec{x})=%
{\displaystyle\sum\limits_{\lambda=1}^{2}}
\int dk^{3}\alpha_{\lambda}(\vec{k})\ \vec{\varepsilon}_{\lambda}(\vec
{k})\left(  \frac{1+i}{2}\text{ }e^{i\vec{x}\cdot\vec{k}}+\frac{1-i}{2}\text{
}e^{-i\vec{x}\cdot\vec{k}}\right)  \label{kjh}%
\end{equation}
We have the same situation for the equation $\partial_{i}\Pi_{i}=0,.$ which
means that \
\begin{equation}
\vec{\Pi}(\vec{x})=%
{\displaystyle\sum\limits_{\lambda=1}^{2}}
\int dk^{3}\beta_{\lambda}(\vec{k})\ \vec{\varepsilon}_{\lambda}(\vec
{k})\left(  \frac{1+i}{2}\text{ }e^{i\vec{x}\cdot\vec{k}}+\frac{1-i}{2}\text{
}e^{-i\vec{x}\cdot\vec{k}}\right)  \label{qsd}%
\end{equation}
where the $\beta_{\lambda}(\vec{k})$ are real components of the vector
$\vec{\beta}(\vec{k}).$ We insert the expressions (\ref{kjh}) and (\ref{qsd})
into the Hamiltonian (\ref{486}) to obtain the following reduced form%
\begin{equation}
H=\frac{(2\pi)^{3}}{2}%
{\displaystyle\sum\limits_{\lambda^{\prime}=1}^{2}}
\int dq^{3}\text{ }\left(  \beta_{\lambda^{\prime}}(\vec{q})^{2}+\left\vert
\vec{q}^{2}\right\vert \text{ }\alpha_{\lambda^{\prime}}(\vec{q})^{2}\right)
\end{equation}
\qquad

At this point let's calculate the bracket $\left\{  \vec{\Lambda}(\vec
{x}),H\right\}  .$ Explicitly,%
\begin{align*}
\left\{  \vec{\Lambda}(\vec{x}),H\right\}   &  =(2\pi)^{3}%
{\displaystyle\sum\limits_{\lambda,\lambda^{\prime}=1}^{2}}
\int dk^{3}dq^{3}\left(  \left\{  \alpha_{\lambda}(\vec{k}),\beta
_{\lambda^{\prime}}(\vec{q})\right\}  \beta_{\lambda^{\prime}}(\vec
{q})+\left\vert \vec{q}^{2}\right\vert \text{ }\left\{  \alpha_{\lambda}%
(\vec{k}),\alpha_{\lambda^{\prime}}(\vec{q})\right\}  \alpha_{\lambda^{\prime
}}(\vec{q})\right) \\
&  \ \ \ \ \ \ \ \ \ \ \ \ \ \ \ \ \ \vec{\varepsilon}_{\lambda}(\vec
{k})\left(  \frac{1+i}{2}\text{ }e^{i\vec{x}\cdot\vec{k}}+\frac{1-i}{2}\text{
}e^{-i\vec{x}\cdot\vec{k}}\right)  .
\end{align*}
According to $(\ref{jij}),$ this equation should be compared with (\ref{qsd})
to immediately get the brackets%
\begin{equation}
\left\{  \alpha_{\lambda}(\vec{k}),\beta_{\lambda^{\prime}}(\vec{q})\right\}
=-\frac{1}{(2\pi)^{3}}\delta_{\lambda\lambda^{\prime}}\delta(\vec{k}-\vec
{q})\text{ \ \ \ \ \ },\text{ \ \ }\left\{  \alpha_{\lambda}(\vec{k}%
),\alpha_{\lambda^{\prime}}(\vec{q})\right\}  =0 \label{951258}%
\end{equation}
We also have
\begin{align*}
\left\{  \vec{\Pi}(\vec{x}),H\right\}   &  =(2\pi)^{3}%
{\displaystyle\sum\limits_{\lambda,\lambda^{\prime}=1}^{2}}
\int dk^{3}dq^{3}\left(  \left\{  \beta_{\lambda}(\vec{k}),\beta
_{\lambda^{\prime}}(\vec{q})\right\}  \beta_{\lambda^{\prime}}(\vec
{q})+\left\vert \vec{q}^{2}\right\vert \text{ }\left\{  \beta_{\lambda}%
(\vec{k}),\alpha_{\lambda^{\prime}}(\vec{q})\right\}  \alpha_{\lambda^{\prime
}}(\vec{q})\right) \\
&  \ \ \ \ \ \ \ \ \ \ \ \ \ \ \ \ \vec{\varepsilon}_{\lambda}(\vec{k})\left(
\frac{1+i}{2}e^{i\vec{x}\cdot\vec{k}}+\frac{1-i}{2}e^{-i\vec{x}\cdot\vec{k}%
}\right)  .
\end{align*}
and from (\ref{kjh})
\[
-\Delta\vec{\Lambda}(\vec{x})=%
{\displaystyle\sum\limits_{\lambda=1}^{2}}
\int dk^{3}\vec{k}^{2}\text{ }\alpha_{\lambda}(\vec{k})\ \vec{\varepsilon
}_{\lambda}(\vec{k})\left(  \frac{1+i}{2}e^{i\vec{x}\cdot\vec{k}}+\frac
{1-i}{2}e^{-i\vec{x}\cdot\vec{k}}\right)  .
\]
With the help of the equation (\ref{ijii}), we identify the brackets
\begin{equation}
\left\{  \beta_{\lambda}(\vec{k}),\alpha_{\lambda^{\prime}}(\vec{q})\right\}
=\frac{1}{(2\pi)^{3}}\delta_{\lambda\lambda^{\prime}}\delta(\vec{k}-\vec
{q})\text{ \ \ \ \ },\text{ \ \ \ }\left\{  \beta_{\lambda}(\vec{k}%
),\beta_{\lambda^{\prime}}(\vec{q})\right\}  =0 \label{mlop}%
\end{equation}
Finally, we deduce the brackets $\{\Lambda^{k}(\vec{x}),\Pi_{j}(\vec{y})\}$,
using the relations (\ref{951258}) and (\ref{mlop}) and closure relation
(\ref{sss})%
\begin{equation}
\{\Lambda_{i}(\vec{x}),\Pi_{j}(\vec{y})\}=-\frac{1}{(2\pi)^{3}}\int
dk^{3}\left(  \delta_{ij}-\frac{k_{i}k_{j}}{\left\vert k^{2}\right\vert
}\right)  e^{i\vec{k}\cdot(\vec{x}-\vec{y})}=-\delta_{ij}^{\bot}(\vec{x}%
-\vec{y})
\end{equation}
This result is identical to the one obtained in the previous section.

\subsection{Conclusion}

\qquad We have successfully studied the free Klein-Gordon and Maxwell fields
in both position and momentum spaces using the generalized CI method. We have
shown through these examples the efficiency of this approach in the treatment
of constrainted non integrable systems.

We first found the relations between the fields and their derivatives at the
initial instant with the help of the first-order Tayor expansion and the
equations of motion. Then, by postulating the Hamilton equations near this
instant, we deduced the Dirac brackets between initial conditions by a direct
identification. Finally, the brackets at any later time were deduced using
their time covariance.

The calculation with the Klein Gordon field was straightforward because this
field is completely independent of its conjugate momentum, unlike the Maxwell
field where the different components are related by the constraints and the
gauge-fixing conditions. This problem was solved in the position space by the
Helmholtz theorem, and in the momentum space, by choosing completely
independent Fourier transforms.

This work shows that the generalized CI approach can give good results in
field theory and thus deserves to be used for the quantization of singular
systems alongside other methods.

\end{document}